# Transcriptomic Models for Immunotherapy Response Prediction Show Limited Cross-cohort Generalisability

Yuheng Liang[1]*, Lucy Chhuo[1,2], Ahmadreza Argha[3], Nona Farbehi[3], Lu Chen[4], Roohallah Alizadehsani[5], Mehdi Hosseinzadeh[6], Min Yang[7], Thantrira Porntaveetusm[8], Youqiong Ye*[9], Hamid Alinejad-Rokny*[1, 10]

[1]*UNSW BioMedical Machine Learning Lab (BML), School of Biomedical Engineering, UNSW Sydney, Sydney, NSW, 2052, Australia*

[2]*Garvan Institute of Meidical Research, Sydney, NSW, 2000, Australia*

[3]*School of Biomedical Engineering, UNSW, Sydney, NSW, 2052, Australia*

[4]*Department of Computer Science, Zhejiang University, Hangzhou, Zhejiang, 310027, China*

[5]*Institute for Intelligent Systems Research and Innovation, Deakin University, Geelong, Victoria, Australia*

[6]*School of Engineering & Technology, Duy Tan University, Da Nang, Vietnam*

[7]Shenzhen Key Laboratory for High Performance Data Mining, Shenzhen Institute of Advanced Technology, Shenzhen, China

[8]*Center of Excellence in Precision Medicine and Digital Health, Faculty of Dentistry, Chulalongkorn University, Bangkok, Thailand*

[9] *Shanghai Institute of Immunology, Shanghai Jiao Tong University School of Medicine, Shanghai 200025, China*

[10]*Visiting Scholar (Collaborative Projects), Center of Excellence in Precision Medicine and Digital Health, Chulalongkorn University, Bangkok, Thailand*

**Email**: h.alinejad@unsw.edu.au (Alinejad-Rokny, H), yuheng.liang@student.unsw.edu.au (Liang, Y), youqiong.ye@shsmu.edu.cn (Ye, Y)

**Running title: Liang Y et al /** Cross-Cohort Evaluation of Transcriptomics Immune Checkpoint Inhibitors Response Prediction Models

**Total Counts of Words:** 8468

**Total Counts of Figures:** 5

**Total Counts of Tables:** 4

**Total Counts of Supplementary Tables:** 5

**Total Counts of Supplementary Figures:** 1

**Total References:**

## Abstract

Immune checkpoint inhibitors (ICIs) have transformed cancer therapy; yet substantial proportion of patients exhibit intrinsic or acquired resistance, making accurate pre-treatment response prediction a critical unmet need. Transcriptomics-based biomarkers derived from bulk and single-cell RNA sequencing (scRNA-seq) offer a promising avenue for capturing tumour-immune interactions, yet the cross-cohort generalisability of existing prediction models remains unclear.We systematically benchmark nine state-of-the-art transcriptomic ICI response predictors, five bulk RNA-seq-based models (COMPASS, IRNet, NetBio, IKCScore, and TNBC-ICI) and four scRNA-seq-based models (PRECISE, DeepGeneX, Tres and scCURE), using publicly available independent datasets unseen during model development. Overall, predictive performance was modest: bulk RNA-seq models performed at or near chance level across most cohorts, while scRNA-seq models showed only marginal improvements. Pathway-level analyses revealed sparse and inconsistent biomarker signals across models. Although scRNA-seq-based predictors converged on immune-related programs such as allograft rejection, bulk RNA-seq-based models exhibited little reproducible overlap. PRECISE and NetBio identified the most coherent immune-related themes, whereas IRNet predominantly captured metabolic pathways weakly aligned with ICI biology. Together, these findings demonstrate the limited cross-cohort robustness and biological consistency of current transcriptomic ICI prediction models, underscoring the need for improved domain adaptation, standardised preprocessing, and biologically grounded model design.

**KEYWORDS:** Immune Checkpoint Inhibitors (ICI), Transcriptomic Biomarkers, Single-cell RNA Sequencing (scRNA-seq), Cross-cohort Benchmarking, Model Generalisation.

## Introduction

Immune checkpoint inhibitors (ICIs) have fundamentally reshaped the landscape of cancer therapy by enabling the patient's immune system to eliminate malignant cells[1]. Agents targeting PD-1, PD-L1, or CTLA-4 have produced durable remissions across multiple tumour types, leading tumour-agnostic regulatory approvals[2-6] (**Figure 1A**). However, only a minority of patients respond to ICIs, with response rates ranging from 0 to 40% across cancer [7]. Resistance arises from both tumour-intrinsic mechanisms, including deficient tumour antigen presentation [8] and dysregulated immune checkpoint-related pathways such as the WNT/β-catenin pathway, mitogen-activated protein kinase (MAPK) and Interferon-gamma (IFN-$\gamma$) pathways[9]. Tumour-extrinsic factors include lack of T-cells, presence of immunosuppressive cells (such as regulatory T cells), and activation of immuno-resistance pathways[10]. Both intrinsic and acquired resistance [11] remain major barriers to effective immunotherapy, highlighting an unmet need for accurate prediction of patient response before treatment initiation in precision immuno-oncology[9].

A broad spectrum of biomarkers has been investigated to address this challenge, as illustrated in **Figure 1B**. Clinical and molecular features such as PD-L1 protein expression[12], tumour mutational burden (TMB)[13], microsatellite instability (MSI)[14], neoantigen load[15], and tumour-infiltrating lymphocyte (TIL) density[16] each capture specific mechanisms that shape immune sensitivity. Nevertheless, these biomarkers exhibit limited sensitivity or specificity to serve as a standalone predictor across cancer types or immune contexts, and discordant cases remain common. For instance, PD-L1-negative tumours may respond to ICIs[17], MSI-high tumours do not always show durable benefit[18], and TMB is confounded by immune-excluded ("cold") tumour microenvironments[19]. These limitations have catalysed interest in transcriptomic biomarkers, which reflect the integrated activity of tumour-intrinsic pathways, immune infiltration, stromal interactions, and cell-cell communication within the tumour microenvironment (TME).

Transcriptomics data, such as bulk RNA sequencing (bulk RNA-seq) and single-cell RNA sequencing (scRNA-seq), capture snapshots of the organism's transcriptome across different dimensions, providing complex information on the regulation of genes and associated pathways. In terms of ICI prediction, samples collected within different tissues, usually Formalin-fixed paraffin-embedded (FFPE) tumour samples, lymph nodes (LNs), or Peripheral Blood Mononuclear Cells (PBMCs), allow comparison of gene expression across different tissues. Samples extracted at time points prior and/or after immunotherapy treatment allow comparison of the TME before and after treatment, providing substantial potential for studying ICI-related biomarkers, thereby promoting the efficiency of immunotherapy[20]. Bulk RNA-seq studies have identified reproducible immune-associated pathways linked to ICI sensitivity, such as IFN-γ signalling[21], cell apoptosis and proliferation processes[22], inflammatory myeloid activation[23], forming signature-based predictors such as the T cell-inflamed gene expression profile (GEP)[24], TIDE (T-cell dysfunction and exclusion model)[25] and IMPRES (immune checkpoint gene-pair signature)[26].

Single-cell RNA sequencing, on the other hand, has revolutionised the understanding of the TME by enabling high-resolution characterisation of immune cell heterogeneity, facilitating the discovery of cell states and interactions[27]. scRNA-seq studies have revealed that ICI responders often harbour progenitor-exhausted CD8 T cells that maintain proliferative potential and clonally expand following PD-1 blockade[28]. Conversely, non-responders frequently display enrichment of immunosuppressive cell populations, such as regulatory T cells (Tregs), dysfunctional tumour-associated macrophages, or myeloid-driven inhibitory programs[28]. In addition, RNA-seq and scRNA-seq often integrated to investigate cell-specific signatures that predict immune oncology landscape, providing a more comprehensive view of transcriptomic signatures[23,29].

Despite these advances, recent transcription-based ICI response predictors (**Figure 2**) face substantial challenges in robustness and generalisability[30]. Existing approaches broadly fall into two conceptual categories. The first comprises fixed-signature or biologically constrained models, such as IKCScore[31], Tres[32], ENLIGHT[33], and TNBC-ICI[34], which rely on predefined gene panels or T-cell-related signatures and apply statistical or machine-learning frameworks for response prediction (**Figure 2A**). These models benefit from biological interpretability and robustness of curated signatures; however, their reliance on predefined gene sets may limit adaptability to diverse tumour contexts and evolving transcriptomic landscapes. Moreover, signature maintenance and updating remain ongoing challenges.

The second category includes machine-learning-driven models that learn predictive features directly from transcriptomic data. These models can be further divided into end-to-end cohort-trained models and domain-adaptive models (Figure **2B**). End-to-end approaches, including NetBio[35], PRECISE[30], scCURE[36], and DeepGeneX[37] typically require retraining on each target cohort. They employ diverse architectures such as logistic regression, XGBoost, multilayer perceptrons (MLPs), and deep neural networks (DNNs). For example,

PRECISE leverages fine-grained single-cell profiles to identify response-associated cell populations using Boruta feature selection and reinforcement learning, followed by LOOCV-based XGBoost prediction. While these models integrate biomarker discovery and prediction within a unified framework, their dependence on cohort-specific retraining limits single-sample deployment and large-scale clinical transferability.

In contrast, domain-adaptive models such as COMPASS[38] and IRNet[39] incorporate pretraining strategies to improve generalisability (**Figure 2C**). COMPASS, for instance, employs a transformer-based architecture pretrained on over 10,000 bulk RNA-seq profiles from The Cancer Genome Atlas (TCGA)[40] (TCGA, https://www.cancer.gov/tcga) to learn general transcriptomic representations. This pretrained foundation is subsequently fine-tuned on multiple ICI-treated cohorts using linear or partial fine-tuning strategies, aiming to enhance cross-cohort transferability.

Nevertheless, building transcriptomics-based machine learning model with robust domain adaptation is challenging. This is because ICI-related transcriptomics cohorts, especially scRNA-seq cohorts are cancer-specific, often imbalanced, and small in size[40-42]. This causes predictors to be biased toward the majority class and increase false negatives. When integrating scRNA-seq datasets, differences in technologies (Smart-seq, 10X genomics) necessitate experiment-specific normalisation, and researchers must consider whether to use batch-aware modelling strategies or batch-effect removal across large cohorts, making methodological choices more complicated. Hence, most scRNA-seq models for ICI response prediction are trained and evaluated on small or technology-specific cohorts, raising concerns about overfitting and model transferability[30,37,39,44]. As a common practice, most models focus on overall accuracy, which can mask imbalance-driven errors. Bulk RNA-seq cohorts are more feasible in terms of sample size and diversity across different cancers. When training with a larger cohort such as TCGA[40], differences in experimental platforms, preprocessing pipelines, normalisation methods, and clinical composition also introduce batch effects and can substantially alter gene expression landscapes. In addition, most prediction models rely solely on sample-level prediction, ignoring cell-type contributions in bulk RNA-seq or collapsing scRNA-seq into pseudo-bulk RNA-seq by averaging gene expression values[35,37,39]. Interpretability also varies substantially across models, with some reporting only gene-level importance and others focusing solely on gene-set-level outputs. To enable fair comparisons, a rigorous, multi-modality benchmark is therefore needed to evaluate how current state-of-the-art models address these challenges.

In this study, we performed a systematic survey of immune checkpoint inhibitor (ICI) response prediction models published up to **20 November 2025** (*See Methods*), we identified **all existing transcriptomics-based models** that explicitly perform ICI response prediction and provide publicly available and reproducible implementations. We evaluate five RNA-seq models (COMPASS[38], NetBio[35], IRNet[39], IKCScore[31], TNBC-ICI[34]) and four single-cell RNA-seq models (PRECISE[30], DeepGeneX[37], Tres[32], scCURE[36]) that were specifically developed for ICI response prediction, representing diverse machine-learning architectures, conceptual frameworks, and biological assumptions. **Table 1** summarises the input modalities, architectural principles, and reported tasks for each model. Through cross-cohort evaluation across independent pan-cancer datasets, we assess how current transcriptomic predictors generalise across cancer types, technologies, and data modalities, and identify convergent biological signals that may represent robust biomarkers. This benchmark aims to guide the development of more reliable and interpretable ICI prediction frameworks and clarify the complementary roles of bulk and single-cell transcriptomics in precision immuno-oncology.

## Results

### Generalisation Performance Across Unseen Bulk and scRNA-seq Cohorts

We evaluated all nine models on six independent, previously unseen cohorts and assessed predictive performance using accuracy, macro F1 score, and area under the receiver operating characteristic curve (AUC). For COMPASS, both linear fine-tuning (LFT) and partial fine-tuning (PFT) variants were evaluated. For IRNet, one dataset provided treatment-specific annotations, enabling separate evaluation on PD-1 and PD-L1 subsets and allowing assessment of drug-conditioned performance. **Table 2** summarises predictive performance across bulk and single-cell RNA-seq datasets, together with class imbalance ratios and runtime, capturing both predictive robustness and practical usability.

Across all datasets, class imbalance was substantial, with imbalance ratios ranging from 0.25 to 0.94, underscoring the importance of macro F1 as a primary performance metric. Overall, all models demonstrated limited robustness when transferred to unseen cohorts. As shown in **Table 2**, predictive performance varied markedly across datasets for all models, indicating strong dependence on cohort-specific transcriptomic context, tumour type, and baseline immune composition. This dataset-dependent behaviour suggests that differences in tumour microenvironmental states and immune infiltration substantially influence model performance.

Bulk RNA-seq-based models (COMPASS, IRNet, NetBio, IKCScore and TNBC-ICI) exhibited pronounced cross-cohort variability and generally limited generalisability. IKCScore, a fixed-signature statistical model, achieved accuracies ranging from 0.39 to 0.69, AUC values from 0.52 to 0.89, and macro F1 scores between 0.48 and 0.62 across cohorts. Although AUC occasionally approached strong discrimination, the corresponding F1 scores remained moderate, suggesting that high ranking ability did not consistently translate into balanced classification performance under class imbalance. This discrepancy highlights the potential sensitivity of fixed-signature models to cohort-specific expression distributions.

TNBC-ICI, developed specifically for triple-negative breast cancer, demonstrated limited transferability beyond its original disease context. Across the evaluated cohorts, accuracies ranged from 0.34 to 0.69, AUC values from 0.48 to 0.52, and macro F1 scores from 0.00 to 0.22. The near-random AUC values and extremely low F1 scores indicate poor discrimination and substantial minority-class prediction failure, reinforcing the challenge of applying tumour-specific gene panels to heterogeneous cross-cancer cohorts.

COMPASS consistently achieved modest performance across datasets, with accuracies ranging from 0.28 to 0.50 and AUC values between 0.31 and 0.55. Partial fine-tuning marginally outperformed linear fine-tuning; however, neither approach achieved reliable discrimination under class imbalance, as reflected by macro F1 scores as low as 0.08. In the Ribas et al. cohort, large discrepancies between accuracy/AUC and F1 highlighted poor minority-class prediction, suggesting limited clinical utility in imbalanced real-world settings.

IRNet demonstrated comparatively more stable performance across two of the three bulk cohorts, with accuracies ranging from 0.32 to 0.69 and AUC values from 0.32 to 0.75. Notably, performance varied noticably when stratified by treatment type: the PD-L1 subset showed improved discrimination relative to the PD-1 subset in the Poddubskaya et al. cohort, indicating sensitivity to drug context. Importantly, IRNet exhibited smaller divergence between accuracy and F1 (F1: 0.44-0.54) than other bulk RNA-seq models, suggesting greater robustness to class imbalance. In contrast, NetBio achieved above-chance performance on two of three cohorts but also produced implausible perfect metrics (accuracy = 1.00, F1 = 1.00, AUC = 1.00) on the Ribas et al.[45]. dataset despite a high imbalance ratio (0.85). Such results strongly suggest cohort-specific overfitting, likely exacerbated by LOOCV-based retraining on small datasets, and highlight the risk of inflated performance estimates in limited-sample settings.

In comparison, scRNA-seq-based models generally achieved higher predictive performance, albeit with substantial dataset dependence. Using T-cell-restricted subsets, PRECISE showed the strongest overall performance, with accuracies ranging from 0.59 to 0.79, AUC values from 0.45 to 0.87, and macro F1 scores from 0.38 to 0.81. Performance varied considerably across cohorts, most notably between Gondal et al. (F1 = 0.38) and Luoma et al. (F1 = 0.81), reflecting differences in cohort composition and immune context. Importantly, feature refinement improved AUC, F1, and accuracy in two of three datasets, indicating that robust feature selection can partially mitigate cross-cohort variability.

scCURE achieved modest but relatively balanced performance, with accuracies ranging from 0.56 to 0.83, AUC values between 0.50 and 0.67, and macro F1 scores from 0.50 to 0.90. While discrimination remained close to chance in terms of AUC, the comparatively stable F1 with the highest value of 0.9 suggest moderate robustness to class imbalance. However, the limited AUC range indicates constrained ranking capacity, implying that predictive signal may be cohort-specific and weakly transferable. Statistical model Tres performed near chance level across all datasets (accuracy: 0.39-0.60; AUC: 0.34-0.66; F1: 0.38-0.58), suggesting limited transferability of resilience-based statistical signatures to unseen scRNA-seq cohorts.

DeepGeneX, which relies on a compact gene panel derived from a small melanoma cohort, showed variable generalisation: predictive performance was retained in datasets with similar immune contexts but deteriorated in cohorts with divergent tumour microenvironments or sequencing technologies. Together, these results underscore the sensitivity of scRNA-seq-based predictors to training cohort composition, cell-state diversity, and technical heterogeneity.

Runtime analysis further revealed clear differences in practical usability. scRNA-seq-based models generally required substantially longer runtimes than bulk RNA-seq approaches, with the exception of Tres, which is a purely statistical method and therefore computationally efficient. scCURE exhibited the longest runtime overall, largely due to its LOOCV feature selection and repeated retraining, which exceeded 280,000 seconds in all datasets. Among bulk RNA-seq models, NetBio incurred the greatest runtime owing to cohort-specific retraining of logistic regression classifiers, whereas IRNet and COMPASS were comparatively efficient when applied in inference mode.

### Cross-dataset comparison

Across bulk RNA-seq cohorts (Cho, Ribas, Poddubskaya), performance patterns were dataset-dependent rather than architecture-dependent. The Cho et al. cohort generally yielded stronger discrimination across several models, whereas the Ribas et al. cohort showed marked instability, with large discrepancies between AUC and macro F1 under high class imbalance (0.85). The Poddubskaya et al. dataset exhibited intermediate performance but revealed treatment-specific effects, as IRNet performed better in the PD-L1 subset than in PD-1. Importantly, no bulk model consistently outperformed others across all three cohorts. Performance variability appeared driven primarily by tumour type, immune context, and transcriptomic distribution rather than modelling strategy, highlighting limited cross-cohort generalisability.

In single-cell datasets (Gondal, Franken, Luoma, Reinstein), overall performance was higher than in bulk cohorts, yet substantial variability persisted. The Reinstein et al. cohort performs consistently stronger compared to other datasets, specifcally regarding F1 value. Considering this cohort is imbalanced (imbalanced rate = 0.26), the high mean F1 (~0.8) indicates the scRNA-seq models learnt the features from the minority class effectively. The Luoma et al. cohort consistently supported stronger discrimination, particularly for PRECISE, whereas Gondal et al. showed lower macro F1 across several models despite moderate AUC values. Franken et al. demonstrated mixed behaviour, with model ranking shifting between cohorts. Notably, no single-cell model maintained dominance across all datasets. These results indicate that although cell-type-resolved modelling improves signal capture, predictive robustness remains highly sensitive to cohort composition, immune-state heterogeneity, and technical differences.

### Pathway-level Biomarker Identification and Clustering

To evaluate the biological relevance and consistency of biomarkers identified by each model, we systematically analysed pathway-level signals and higher-order biological concepts associated with immune checkpoint inhibitors response. Rather than focusing on individual genes alone, we sought to determine whether different modelling strategies converge on shared immune-oncology-relevant programs or recover distinct, model-specific biological themes.

Models were grouped into two categories: (i) models relying on fixed gene panels (IKCScore, DeepGeneX, and TNBC-ICI) and (ii) adaptive models that generate cohort-dependent biomarker sets (COMPASS, IRNet, NetBio, PRECISE and Tres). For each group, enriched pathways were visualised using the EnrichmentMap workflow in Cytoscape, followed by AutoAnnotate to generate higher-order biological concepts. Among fixed-panel models, substantial heterogeneity was observed in the number of significant pathways identified (IKCScore: 63; DeepGeneX: 21 and TNBC-ICI: 95), reflecting differences in gene panel size and pathway enrichment breadth. Clustering of these pathways yielded 15 overarching biological concepts (**Figure 4A-C**).

Despite sparse gene panels, several immune-related programs were recurrent across models. IKCScore, composed of positively associated immune checkpoint genes, immune score components, and a negatively associated keratinisation-associated (KRT) score, predominantly recovered pathways related to *immune system activation*, *adaptive immunity*, and *T-cell receptor signalling*. These themes overlapped substantially with TNBC-ICI, which also enriched for immune system, adaptive immune response, cell adhesion molecule interactions, and *TGF-β-related signalling*. KRT score outlines only 2 biological pathways independent of other clusters, *ion channel transport* and *pre-implantation embryo*. Ion channel transportations are essential processes of keratinisation[46], and keratinisation process is crucial in the development of pre-implantation embryo[47]. A total of 16 pathways and six biological concepts were shared between IKCScore and TNBC-ICI, suggesting that independently curated immune-focused gene panels converge on core adaptive immune programs relevant to ICI response. DeepGeneX, derived from a compact melanoma-specific gene signature, identified several immune-associated pathways, including *T-cell modulation* and *IGF-insulin-IGFBP signalling*. However, due to the small size of its gene set, most enriched pathways were supported by only one or two genes, limiting interpretability and reducing robustness across cohorts.

Across fixed-panel models, four pathways including *autoimmune thyroid disease*, *B-cell activation*, *allograft rejection*, and *natural killer cell interactions* were consistently enriched, forming four shared biological concepts. These pathways converged on a shared cytotoxic effector axis centred on Granzyme B (*GZMB*), a key molecule produced by cytotoxic $CD8^+$ T cells and natural killer (NK) cells that induces apoptosis in target cells following immune activation[48]. Members of the granzyme family, particularly *GZMB*, have been shown to predict response to immune checkpoint blockade in cutaneous melanoma, especially in patients treated with anti-PD-1/PD-L1 therapies[49]. These recurrent themes highlight convergence on adaptive immune activation and cytotoxic effector programs, supporting their central role in ICI response biology.

On the other hand, the underlying bio-mechanisms of the important features generated by the testing cohorts are organised into 28 functional biological concepts, capturing higher-order immune, stromal, and metabolic processes (**Figure 4D-F**). We observed marked heterogeneity in the number of significant pathways identified: COMPASS recovered 15 pathways, IRNet 23, NetBio 38, PRECISE 239, and Tres 331 pathways (**Figure 4E**). Here, scRNA-seq models output more biological pathways compared to bulk RNA-seq models. This wide range reflects fundamental differences in model architectures, feature selection strategies, and the granularity at which biological signals are extracted.

Across all models, the overall pathway intersection structure was highly sparse, indicating limited reproducibility of specific pathway-level biomarkers. As shown in **Figure 3D** and **3E**, most pathways clustered within individual models, with minimal overlap across approaches. Notably, single-cell RNA-seq-based models exhibited modest convergence, whereas bulk RNA-seq-based models showed no shared pathways within modality. Among scRNA-seq models, four immune disease-related pathways, *Allograft rejection*, *Graft-versus-host disease*, *Type I diabetes mellitus*, and *Autoimmune thyroid disease* were consistently identified. The recurrent identification of *GZMB*-centred pathways across independent scRNA-seq models underscores its role as a robust, cell-intrinsic marker of effective anti-tumour immunity.

Beyond individual pathways, several higher-order biological themes were shared across models. Although the bulk RNA-seq models did not converge on identical pathway terms, broader concepts overlapped between bulk and single-cell approaches. For example, *eukaryotic translation termination* identified by COMPASS, PRECISE, NetBio, and Tres, may influence immune-checkpoint inhibitor (ICI) response through its role in nonsense-mediated decay (NMD) and the regulation of aberrant peptide production. Reduced NMD activity can increase tumour immunogenicity by allowing the accumulation of mutation-derived peptides that may act as neoantigens, thereby enhancing responsiveness to immunotherapy[50]. In addition, shared concepts such as *cell signalling responses* and *transcription foxo checkpoints* and *interferon gamma response* shared concepts were primarily driven by COMPASS, PRECISE, and Tres, primarily identified by COMPASS, PRECISE, and Tres, suggest partial cross-modality convergence at the level of T-cell activation and survival. These concepts reflect coordinated immune activation, cytotoxic effector function, and metabolic reprogramming, hallmarks of responsive tumour-immune microenvironments.

Among all models, PRECISE demonstrated the greatest degree of biological convergence, sharing pathways or concepts with every model except IRNet. PRECISE showed particularly strong overlap with Tres, sharing 54 biological pathways including immune-related *innate immune system*, *Interleukin-4 and Interleukin-13 signalling*, and *Interferon-γ signalling*. At concept level, interestingly PRECISE does not generate any unique biological concepts, highlighting its convergence. PRECISE shown 11 overlapped concepts with Tres, outlining immune-related biological concepts *downstream cell bcr,* reflecting activation and differentiation of tumour-infiltrating B cells. B cells play a complementary role on immunotherapy responses through interacting with T cells to promote immune activation, enhance antitumor immune priming and sustain T-cell–mediated responses following checkpoint blockade[51]. Together, these findings highlight PRECISE's ability to capture biologically coherent, immune-relevant programs that generalise across datasets and modelling frameworks.

Tres identified the largest number of pathways, suggesting reduced specificity or overly permissive enrichment. While Tres shared immune-related concepts with COMPASS, including *egfr g2 phases*. Aberrant EGFR activation can promote immune evasion by reducing antigen presentation and suppressing cytotoxic T-cell infiltration, while also driving rapid cell-cycle progression through the G2/M checkpoint. Consequently, tumours with strong EGFR-driven cell-cycle activity are often linked to reduced responsiveness to ICIs, whereas inhibition of EGFR signalling may enhance tumour immunogenicity and improve immunotherapy efficacy[52]. Besides, Tres also recovered a broad set of pathways with limited relevance to immune checkpoint biology. These included concepts related to *Golgi-associated ferroptosis*, *aerobic glycolysis*, and *pre-mRNA splicing*, which, although relevant to cancer biology and therapeutic targeting, are not directly linked to immunotherapy[53]. The breadth and heterogeneity of pathways identified by Tres suggest that its resilience-based signatures may lack sufficient discriminative specificity in the context of ICB response prediction.

NetBio did not share exact pathway terms with other models but showed overlap at the level of broader biological concepts, particularly with Tres, COMPASS, and PRECISE. This suggests that while NetBio may capture high-level biological themes, it lacks reproducible, fine-grained pathway signals. One notable shared concept between NetBio and COMPASS was *collagen-focal adhesion interactions*, which are increasingly recognised as modulators of immune infiltration and therapeutic response. Targeting collagen-rich tumour stroma has been proposed as a strategy to enhance immunotherapy efficacy by improving immune cell access and function[54]. NetBio also uniquely identified cancer-associated pathways such as *Liver cancer progression* and *Holleman vincristine resistance*, reflecting its broader oncogenic focus rather than immune specificity.

COMPASS shared a limited number of pathways with PRECISE and Tres, including several related to infectious disease processes, such as viral infection pathways *Human papillomavirus infection*. While these pathways are not cancer-specific, they reflect conserved innate immune activation programs. Viral and tumour immunity share overlapping signalling cascades, including interferon responses and antigen presentation machinery, which may explain this partial convergence[55].

Finally, IRNet emerged as a clear outlier in pathway-level analysis. It did not overlap with any other model at either the pathway or concept level and predominantly recovered metabolic pathways, including pyruvate, nitrogen, propanoate, vitamin, and pyrimidine metabolism, as well as RNA degradation and protein digestion. These pathways formed an isolated cluster within the EnrichmentMap network, indicating a strong bias toward cellular metabolism. Although metabolic reprogramming is increasingly recognised as a regulator of immune cell function and immunotherapy resistance[56], the lack of overlap with immune-specific pathways suggests that IRNet may prioritise metabolic features at the expense of canonical immune-oncology programs. Whether these signals reflect true biological mechanisms of response or model-specific artefacts warrants further experimental validation.

EnrichmentMap furthermore revealed notable differences in the pathway-level convergence across datasets, and between the sequencing modalities. Bulk RNA-seq cohorts (Cho et al., Ribas et al., and Poddubskaya et al.), analysed using NetBio, IRNet and COMPASS, exhibited sparse and fragmented enrichment networks, characterised by minimal pathway overlaps, small weakly connected clusters, and numerous isolated pathways. In the Cho et al. (**Figure 5A**) dataset, only a single biological concept, *Rho GTPases activate,* was converged. The pathway is known to regulate innate immunity through acting as molecular switches during immune cell migrations[57]. The Poddubskaya et al. (**Figure 5C**) dataset only identified three converged biological concepts, including signalling infection pathways and valine leucine isoleucine metabolism, which correlate with immunity. Signalling infection pathways include Toll-like Receptors (TLRs) and MAPK pathways initiated by pathogen recognition receptors (PRRs), which regulate innate immune signalling and trigger immune responses[58]. Branched-chain amino acids such as valine, leucine and isoleucine are essential in immune metabolism, helping to activate adaptive immune responses by regulating the function of regulatory T cells (Treg) via mTOR signalling[59]. The Ribas et al. dataset displayed moderately higher connectivity, with five distinct clusters. These included immune-related clusters such as RNA-binding protein *auf1 hnrnp d0* which has been described as an intracellular checkpoint inhibitor[60], as well as the oncogene-associated concept *lui pax8 fusion*. Additional clusters identified were related to metabolic and infection-associated signalling. Overall, bulk RNA-seq-based predictions showed low concordance at the pathway level across unseen cohorts, yet models identified immune-related concepts showing some degree of biological interpretation.

In contrast, scRNA-seq datasets, Gondal et al., Franken et al., Luoma et al and Reinstein et al. analysed utilising DeepGeneX, Tres and PRECISE, demonstrated denser and more highly connected enrichment networks with prominent immune-related pathway clusters. Overall, Gondal et al. showed no overlapping biological concepts across models (**Figure 5D**), whereas Franken et al. exhibited six clusters shared in the significant pathways identified by scRNA-seq based models (**Figure 5E**), Luoma et al identified three overlapping concepts (**Figure 5F**), and Reinstein et al (**Figure 5G**) identified the largest number, 12 overlapping concepts. These networks included *selective chemokine polarization,* highlighting the role of chemokine, which are described as turning 'cold' tumours, i.e., tumours that lack effector T cells or are enriched in Treg cells, into 'hot' tumours by enhancing anti-tumour immunity[61]; *igf insulin growth factor* signalling, highlighting the function of *IG-1FR* gene, which is an important ICI biomarker in cancer treatment; and *antigen signalling via the tcr*, which is the crucial first step for adaptive immunity, and plays a key role in functioning and proliferation of T-cells[62]; immune activation or cytokine cluster involving *signaling interleukin cytokine*, *cancer immunotherapy PD* and *antigen presentation*, strongly associating with effective ICI response, as they indicate an inflamed tumour microenvironment with active immune recognition[63]. Other clusters identified included immunity-related concepts such as *sars cov infection*, *NK cell signalling* and *allograft rejection*, forming large, cohesive modules.

Collectively, these analyses demonstrate that although most models capture biologically meaningful signals, their ability to identify robust and reproducible immune-related pathways varies substantially across architectures. Among all approaches, PRECISE exhibited the highest degree of biological convergence and immune relevance. These findings emphasise the importance of model-agnostic, concept-level analyses for evaluating biomarker robustness and highlight the challenges of translating pathway-level signatures into generalisable predictors of immune checkpoint blockade response. When considered across datasets, the increased connectivity suggests greater functional agreement among scRNA-seq-based models compared to bulk RNA-seq based models. Despite this improved coherence, pathway composition and cluster prominence varied across scRNA-seq cohorts, indicating that the enriched biological themes remain partially dataset-dependent.

## Discussion

This study provides a systematic cross-cohort benchmark of transcriptomics-based ICI response prediction models across bulk and single-cell modalities. When evaluated on strictly unseen datasets processed through a unified pipeline, all models demonstrated reduced performance compared with their originally reported results, highlighting substantial domain-transfer limitations.

Among bulk RNA-seq models, generalisability was consistently limited. COMPASS, despite large-scale pretraining on TCGA and fine-tuning on multiple ICI cohorts, did not demonstrate superior cross-cohort robustness. End-to-end models such as IRNet and NetBio occasionally achieved higher performance; however, these gains were inconsistent and highly dataset dependent. Notably, NetBio produced perfect metrics on one cohort under high imbalance, strongly suggesting overfitting related to LOOCV-based retraining on small datasets rather than true biological generalisation.

Fixed-panel models further illustrated transfer constraints. IKCScore showed moderate but variable discrimination across cohorts, with discrepancies between AUC and macro F1 indicating sensitivity to class imbalance and cohort-specific expression distributions. TNBC-ICI, developed for a specific breast cancer context, performed near chance on heterogeneous datasets, reinforcing the limited portability of tumour-specific gene panels. Collectively, these results indicate that modelling strategy alone does not determine robustness; cohort heterogeneity and immune-context variability are dominant drivers of performance instability in bulk transcriptomic prediction.

Interestingly, the end-to-end bulk RNA-seq based models IRNet and NetBio outperformed the foundation model COMPASS on the unseen cohorts, despite COMPASS being pretrained on more than 10,000 TCGA[40] samples and fine-tuned on 15 independent ICI-treated cohorts. This finding highlights an important limitation of large-scale foundation models in this setting and underscores the challenges of domain transfer in transcriptomic ICI prediction. End-to-end models directly optimise for the prediction task under a single objective and distribution, whereas COMPASS relies on transfer learning. When applied to cohorts whose characteristics diverge from TCGA[40] distributions (e.g., different immune compositions or tumour microenvironments), the fine-tuning may be insufficient for complete domain adaptation. This is a widely recognised challenge of foundation models in precision oncology: broad representation learning does not always translate into optimal performance on narrowly defined clinical tasks.

Single-cell RNA-seq models exhibited comparatively stronger predictive signals but remained highly dataset-dependent. PRECISE achieved the most consistent performance across cohorts and demonstrated improved discrimination following feature refinement, suggesting that cell-state-aware feature selection captures biologically meaningful immune programs. scCURE achieved modest but relatively balanced F1 scores despite near-chance AUC values, indicating limited ranking capacity but some robustness to class imbalance. DeepGeneX generalised partially but deteriorated in cohorts with divergent immune landscapes, likely reflecting overfitting to its small melanoma-derived gene panel. Tres performed at or near chance across unseen datasets, suggesting that resilience-based transcriptional signatures derived from specific sequencing platforms do not readily transfer across technologies or immune contexts. Together, these findings suggest that while cell-type-resolved modelling enhances biological resolution, it does not eliminate cross-cohort domain shift.

Overall, these findings suggest that current bulk RNA-seq prediction models have limited ability to generalise across unseen datasets, especially when tumour type, technical processing, or immune context differ from those present during training. Despite their architectural sophistication, their robustness across cohorts remains incomplete. Although scRNA-seq models exhibit encouraging signals of predictability, their generalisation across unseen datasets remains limited, highlighting the sensitivity of single-cell predictors to dataset size, technology, and cell-state composition.

Pathway-level analysis revealed partial biological convergence across models. Fixed immune-focused panels (IKCScore, TNBC-ICI, DeepGeneX) consistently enriched adaptive immune activation, T-cell receptor signalling, allograft rejection, and natural killer cell pathways, underscoring the centrality of cytotoxic and adaptive immunity in ICI response. PRECISE showed the strongest overall biological coherence, overlapping immune-related concepts with multiple other models. In contrast, Tres generated numerous but heterogeneous pathways with reduced specificity, and IRNet produced largely non-overlapping pathway outputs, likely reflecting architectural bias toward KEGG-restricted metabolic graphs. These differences highlight how model architecture and pathway database choice shape downstream biological interpretation.

Importantly, several enriched pathways reflected general immune activation rather than tumour-specific anti-tumour immunity, underscoring a core challenge in biomarker discovery: distinguishing immune tone from effective anti-cancer immune response. This may partially explain why models trained on immune-activation signatures fail to generalise consistently across tumour types.

Technical factors further influenced performance. Models differed substantially in normalisation assumptions (TPM, TMM, log-TPM, UMI-based scaling), and applying inappropriate preprocessing, such as TPM normalisation to UMI-based data, likely introduced gene-length bias and degraded performance. Batch correction and feature selection strategies also materially affected outcomes, suggesting that normalisation-agnostic or preprocessing-aware architectures may be necessary for reliable deployment.

Practical usability varied markedly. Pretrained models such as COMPASS and IRNet allow single-sample inference and are more compatible with clinical deployment, whereas NetBio, PRECISE, and scCURE require cohort-level retraining, limiting real-time applicability. Runtime analysis further revealed substantial computational disparities, with scCURE and DeepGeneX incurring particularly high computational costs due to repeated retraining and feature elimination procedures. These trade-offs between interpretability, adaptability, and computational burden are non-trivial considerations for translational use.

Overall, our benchmark demonstrates that current transcriptomic ICI prediction models capture biologically meaningful signals but lack consistent robustness when transferred across tumour types, sequencing technologies, and immune contexts. Future models must incorporate stronger domain adaptation strategies, harmonised preprocessing pipelines, and cross-cohort training frameworks to achieve clinically reliable generalisation in precision immuno-oncology.The size of the gene lists produced by different models varied dramatically, which directly affected both the quantity and interpretability of downstream pathways. For example, DeepGeneX produces an extremely concise signature (six genes in the original publication), which necessarily limits pathway discovery to one or two overlapping terms per pathway. In contrast, models like Tres output nearly one hundred genes, producing hundreds of enriched pathways and dense clusters in EnrichmentMap. These inconsistencies in gene-list length reflect fundamental architectural differences, but they also highlight a methodological challenge: short signatures risk missing higher-order biological context, while very large signatures can inflate pathway count without improving specificity. Standardising the granularity of gene outputs or applying post-hoc harmonisation steps may therefore be necessary for fair comparison and biologically consistent biomarker interpretation.

Overall, our benchmarking results reveal that current transcriptomic ICI prediction models exhibit limited robustness when transferred to truly unseen datasets. While each model captures biologically meaningful signal in certain contexts, none generalise consistently across tumour types, sequencing technologies, or immune compositions. These findings underscore a critical gap in current computational immuno-oncology: models must be designed with stronger domain adaptation, more robust cross-cohort training strategies, and normalisation-agnostic architectures to achieve reliable performance in real-world clinical settings.

## Limitations and Future Directions

This study has several important limitations that should be considered when interpreting the results. First, achieving complete methodological fairness is inherently challenging because the benchmarked models span fundamentally different architectural paradigms, including statistical signature-based approaches, end-to-end deep learning models, and foundation-model-based frameworks. These differences affect multiple stages of the analytical pipeline, from feature representation and learning objectives to pathway attribution strategies, rendering direct, one-to-one comparisons imperfect by design. Second, the scRNA-seq component of our benchmark was restricted to T-cell-focused datasets with relatively small sample sizes. While this choice reflects realistic clinical constraints, it inevitably limits model stability and statistical power. This effect is evident in the reduced performance of PRECISE in our benchmark (AUC ≈ 0.6) compared with its originally reported performance on the full Gondal et al. cohort (AUC ≈ 0.8). Small cohort sizes also constrain downstream biological interpretation, particularly for pathway-level analyses that rely on sufficient representation of diverse immune cell states. Third, the size of gene signatures produced by different models varied substantially, ranging from highly compact panels (six genes in DeepGeneX) to large gene sets approaching one hundred genes (Tres). This heterogeneity directly influences the scope and interpretability of pathway enrichment analyses. Extremely small signatures may fail to capture higher-order biological programs, whereas large signatures can inflate the number of enriched pathways without necessarily improving specificity or reproducibility.

These disparities reflect fundamental architectural choices but also highlight the need for harmonised output representations when comparing biological signals across models.

Looking forward, several methodological directions may help overcome the limitations identified in this benchmark. One promising avenue is the use of heterogeneous graph-based models to represent the complex, multi-scale structure of tumour-immune interactions underlying immune checkpoint blockade response. Unlike homogeneous gene-gene or pathway-level graphs, heterogeneous graphs can explicitly encode multiple biological entities, such as genes, immune cell types, tumour cells, pathways, ligands-receptors, and clinical variables, as distinct node types, with typed edges representing regulatory, signalling, spatial, or functional relationships. Such representations naturally align with the biology of immunotherapy, where response emerges from coordinated interactions across molecular, cellular, and tissue scales rather than from isolated gene expression patterns. By jointly modelling tumour-intrinsic programs, immune cell states, and microenvironmental context within a unified graph framework, heterogeneous graph neural networks may improve both cross-cohort robustness and biological interpretability of ICB predictors.

In parallel, large language models (LLMs) offer a complementary opportunity to address limitations in both generalisation and interpretability. LLMs trained or adapted for biomedical domains can act as knowledge-integrating components that reason over structured representations (e.g., pathways, cell-cell interaction graphs, immune ontologies) and unstructured prior knowledge from the literature. Rather than serving as black-box predictors, LLMs could be used to contextualise transcriptomic signals, enforce biologically plausible reasoning, or generate mechanistic explanations linking model predictions to known immunological processes. In the context of ICB prediction, LLMs may help bridge cohort-specific transcriptomic patterns with general immunological principles, mitigating overfitting to dataset-specific artefacts and improving robustness under domain shift.

Importantly, future architectures may benefit from hybrid frameworks that combine heterogeneous graph representations with LLM-based reasoning modules. In such systems, graph-based encoders could learn data-driven representations of tumour-immune states, while LLMs operate at a higher semantic level to integrate prior knowledge, guide feature selection, or interpret model outputs in clinically meaningful terms. This paradigm may be particularly valuable for translating complex multi-omics and single-cell data into actionable insights for precision immunotherapy.

Finally, integrating transcriptomic data with complementary modalities, including clinical phenotypes, tumour burden, blood-based immune markers, spatial tumour microenvironment features, and genomic alterations such as copy number variation, whole-exome sequencing, or whole-genome sequencing, remains a critical direction for improving generalisability. Models that are explicitly designed to accommodate heterogeneous inputs and variable data availability are more likely to perform reliably in real-world clinical settings. Together, these directions point toward a next generation of ICB prediction models that move beyond single-modality, cohort-specific predictors toward biologically grounded, knowledge-aware, and domain-resilient systems capable of supporting robust and interpretable precision immunotherapy.

## Conclusion

Overall, our findings demonstrate that the current models show some predictability on the unseen samples, but overall, the performance in both prediction and biomarker identification is not sufficient. No single model provides a complete or universally stable biomarker set, essential for reliable biological interpretation. Future directions include standardising output formats and benchmarking frameworks that account for model architecture and dataset size. Importantly, our results reinforce the value of single-cell-informed signatures and biologically structured models in revealing reproducible immune pathways relevant for ICI response. As more diverse datasets and multimodal foundation models emerge, future work should prioritise harmonised evaluation criteria and biologically grounded interpretability to accelerate the development of clinically actionable biomarkers.

## Method and Materials

### Model Selection

To construct a balanced and modality-specific benchmarking framework, we conducted a structured survey of immune checkpoint inhibitor (ICI) response prediction models published within the last three years (2022-2025 Nov.). Candidate methods were screened according to predefined inclusion criteria. First, models were required to explicitly perform sample-level ICI response prediction, defined as outputting a binary classification or a continuous probability score corresponding to the response status. Methods developed solely for biomarker discovery, descriptive analysis, or tumour microenvironment profiling were excluded. Second, models had to use transcriptomic data (bulk RNA-seq or single-cell RNA-seq) as their primary input modality; multimodal or multi-omics frameworks were excluded to ensure comparability.

Third, each model needed to provide a reproducible computational pipeline with publicly available code or parameter descriptions sufficient for external implementation.

Following this screening, nine models met all criteria and were selected for benchmarking: five bulk RNA-seq-based models (IRNet[39], NetBio[35], COMPASS[38], IKCScore[31], and TNBC-ICI[34]) and four single-cell RNA-seq-based models (DeepGeneX[37], PRECISE[30], Tres[32], and scCURE[36]). For the purposes of evaluation, models were grouped by input modalities. A model inclusion-exclusion justification is presented in Supplementary Table S1.

**Dataset collection and QC**

To ensure modality-appropriate and unbiased model evaluation, we first conducted a systematic comparison of the datasets used in the original training and validation of each model. Bulk RNA-seq-based models were evaluated exclusively on independent bulk RNA-seq datasets that were not used during their development, while scRNA-seq-based models were evaluated on independent scRNA-seq datasets entirely unseen in their respective training pipelines. A detailed comparison of dataset origin, sample size, cancer type, and sequencing platform is provided in Supplementary Table S2.

Most benchmarking datasets consisted of tumour-derived transcriptomes; however, one scRNA-seq dataset contained PBMC samples due to limited tumour-infiltrating immune cells in the corresponding study. To maintain clinical relevance and avoid confounding from treatment-induced transcriptional changes, only pre-treatment samples were included in all analyses. This ensures that model predictions reflect baseline transcriptomic states and aligns with the intended use of ICI response predictors for prospective decision-making.

**Bulk RNA-seq Datasets**

Three independent bulk RNA-seq datasets not used in any model's original training were selected for benchmarking: Cho et al. (2020)[64], Ribas et al. (2020)[45], and Poddubskaya et al. (2024)[65]. All datasets were obtained from the Gene Expression Omnibus (GEO)[66]. Although the Ribas et al. dataset provides normalised counts in GEO, we additionally retrieved the corresponding raw FASTQ-level counts from the NCBI Sequence Read Archive to ensure consistent processing across cohorts.

Each dataset was processed using a unified pipeline. Only pre-treatment tumour samples were retained. Gene identifiers were first harmonised by mapping to Ensembl gene IDs, followed by annotation to GENCODE v36[67] to ensure consistent gene symbol representation across datasets. Clinical response labels were assigned according to standard RECIST[68] criteria: samples with complete or partial response (CR/PR) were classified as responders, and those with stable or progressive disease (SD/PD) were classified as non-responders. Samples lacking definitive response annotations were excluded. A detailed summary of cancer type, sample size, sequencing platform, and response distribution for all three datasets is provided in **Table 3**.

**Single-cell RNA-seq Datasets**

To evaluate scRNA-seq-based predictors under both integrated and cohort-specific conditions, we selected four independent single-cell RNA-seq datasets, comprising one large integrated dataset (Gondal et al., 2025[69]) and three smaller standalone cohorts (Franken et al., 2024[70] and Luoma et al., 2022[71], Reinstein et al., 2025 [72]). This design allows assessment of model performance in a heterogeneous, batch-corrected multi-study setting and more focused datasets with consistent experimental conditions. For the Gondal et al. dataset, we removed any constituent datasets that were used in the training of existing scRNA-seq prediction models to avoid information leakage. The remaining datasets were merged into a single integrated profile. All scRNA-seq datasets were processed using a standard pipeline: log-1p normalisation (log1p transformation), removal of cells expressing fewer than 100 genes and removal of genes detected in fewer than 3 cells, followed by batch correction using BBKNN for integrated analyses. Gene identifiers were harmonised by mapping to Ensembl IDs and subsequently annotated using GENCODE v36[67], ensuring compatibility across datasets. For Reinstein et al. dataset, since the cell type information is not provided, we carried out a standard cell type annotation process and identified clusters that with high T-cell cluster biomarkers (gene list) as our targeted cells. The full annotation is provided in the Supplementary Figure S1.

Response labels were assigned according to the criteria used in the respective original publications. Reinstein et al. (2025) patients outlined patient responses based on RECIST[68], with CR/PR as responders and PD/SD as non-responders. Luoma et al. (2022) reported limited specificity of RECIST[68] calls in melanoma, we followed published criteria for melanoma outcome assessment[73] and defined responders as patients exhibiting either a volumetric response or ≤10% viable tumour following treatment. Immune response of Franken et al. (2024) was defined by T-cell clonal expansion; Expansion (E) was assigned as responder and Non-Expansion (NE) as non-responder. Gondal et al. (2025) patients categorised as favourable or unfavourable response in the original study were mapped to responders and non-responders, respectively.

**Model-Specific Preprocessing for Consistent and Fair Evaluation**

To ensure fair comparison across models, each dataset was processed using the pre-processing strategy recommended or required by the corresponding original publication. This approach prevents performance differences from arising due to inconsistent normalisation or feature handling.

*COMPASS*

COMPASS requires inputs restricted to the gene universe used in its foundation model. Therefore, we intersected each bulk RNA-seq dataset with the COMPASS gene set and applied TPM normalization followed by log1p transformation, consistent with the original fine-tuning procedure. No additional gene filtering was performed.

*IRNet*

Because IRNet extracts pathway structure from KEGG and expects the full transcriptome input, no gene subsets were applied. Expression matrices were converted to log-TPM values, as described in the original study, to ensure compatibility with the graph neural network architecture.

*NetBio*

NetBio was evaluated using TMM normalisation, matching its initial implementation and ensuring that ssGSEA-derived pathway scores reflect comparable library-size-adjusted counts across samples.

*IKCScore*

IKCScore utilised TPM normalisation on the datasets, ensuring a matched modality between the original implementation and generalised running.

*TNBC-ICI*

TNBC-ICI utilised z-score normalisation of raw data, we utilised raw data as the input, and predict using the RF models built from the training data and fixed gene panel.

*PRECISE*

To minimise noise from confounding or uninformative gene classes, we followed PRECISE preprocessing guidelines by removing non-coding genes, ribosomal protein genes (RPS/RPL/MRP/MTRNR), mitochondrial genes, and genes expressed in <3% of cells. Cells were annotated by response status, timepoint (pre- vs post-treatment), and tissue source; samples with inconsistent or ambiguous response labels were excluded.

*DeepGeneX*

Consistent with its original design, DeepGeneX was applied to patient-level pseudo-bulk profiles, obtained by averaging gene expression across all cells belonging to a given patient prior to quality filters or feature selection.

*Tres*

Tres originally employs TPM normalisation for Smart-seq-based full-length scRNA-seq data to correct for gene-length bias. Because our unseen datasets were generated using 10x Genomics UMI technology, where counts are inherently length-independent, TPM would artificially reintroduce gene-length bias. Accordingly, we applied Counts-per-Million (CPM) normalization, providing a more appropriate basis for scoring T-cell resilience signatures.

*scCURE*

scCURE applies ICI prediction via training the model with both normalised pre-treatment and post-treatment datasets and utilised LOOCV to test on post-treatment samples, hence, for this model only, we included post-treatment data for all scRNA-seq datasets.

**Benchmarking Strategy and Performance Evaluation**

We benchmarked each model by applying unseen bulk RNA-seq datasets to the bulk-based models (IRnet, NetBio, COMPASS, TNBC-ICI, IKCScore) and unseen scRNA-seq datasets to the single-cell-based models (PRECISE, DeepGeneX, scCURE, Tres). Because the models differ in architecture and training dependencies, we applied model-specific validation strategies to ensure fairness.

Statistical model IKCScore and Tres were run accordingly on the unseen datasets. COMPASS provides two fine-tuned versions utilised all provided datasets in the publication: a linear-probing model (LFT) and a partially fine-tuned model (PFT). Both of which were applied directly to the unseen cohorts. IRnet offers pre-trained pathway-aware weights, enabling direct inference without retraining. For Poddubskaya et al., since drug information is provided, we ran predictions on PD-L1 subset (6 samples) and PD-1 subset (54 samples) for testing the predictability of drugs. DeepGeneX supplies a refined gene list derived from recursive feature elimination; we used this gene set to evaluate predictive performance on unseen datasets. Tres provides T cell-resilience gene signatures derived from training cohorts, allowing direct application to new samples. In contrast, NetBio, PRECISE, TNBC-ICI and scCURE require cohort-specific training. For these models, we reproduced the feature extraction and prediction pipelines described in the original publications. Across the models, PRECISE additionally incorporates a two-stage process: (i) initial prediction using all genes, and (ii) Boruta-based feature refinement followed by a second prediction step.

Prediction performance was assessed using accuracy, F1-score, and area under the ROC curve (AUC). For completeness, we also recorded approximate runtime and workflow complexity for each model. All models were evaluated using default hyperparameters without further tuning.

**Gene-set Discovery and Pathways Network**

We benchmarked the biomarkers identified by each of the models to identify if they have identified immune oncology or cancer-related biological concepts. Since the biomarker discovery process between different models are different, we firstly aligned the outputs of each of models to biological pathways doing geneset enrichment analysis (GSEA) on the models. We built the strategy based on the feature and task for each model. scCURE predicts ICI response based on cell cluster changes, hence is excluded from the gene set-analysis. For IRnet and NetBio, since they output importance matrices between pathways and samples directly, we performed a z-score normalisation on the mean importance value across each pathway. To ensure a modest number of pathways, pathways with z-score >=2 are selected as important pathways. The complete list of important biological pathways for IRNet and NetBio are depicted in the Supplementary Table S3.

For the remaining models (COMPASS, Tres, PRECISE, IKCScore, TNBC-ICI and DeepGeneX), which output gene-level importance, we derived methods for extracting model-specific biological pathways. Tres utilised Pearson's correlation between the Tres gene score vector with the patient's expression vector, as a result we derived Expression×Tres (gene_score) to further differentiate the gene feature that contribute to the responders. COMPASS generated importance matrix of gene versus samples. A z-score normalization was proceeded on the mean importance values across Tres and COMPASS samples. To maintain the modest number of genes, features with z-score >= 2.5 are considered as important features. Important features generated by PRECISE were select through the inbuilt feature selector by Boruta Analyzer as the genes confirmed to be important. Although DeepGeneX provides a feature elimination process, considering for benchmarking we only tested the genes provided in the original paper, we utilised the original gene list as the important features. IKCScore and TNBC-ICI also utilised fixed gene panels, hence the gene panel were used as the important features. Specifically, IKCScore utilised three gene lists forming its basis of prediction: Immune score, immune checkpoint and KRT score. Here, each gene list was considered as separate inputs. We then utilise EnrichR[74] to discover the important biological pathways through gene enrichment. We selected biological databases including WikiPathways[75], KEGG[76], Reactome[77] and MSigdb Hallmark[78] to align with IRnet and NetBio outputs. We considered p-value < 0.05 as important pathways. The complete list of important biological pathways for COMPASS, TNBC-ICI, IKCScore, Tres, PRECISE and DeepGeneX are demonstrated in the Supplementary Table S4.

We separated the models into two categories: models that utilised fixed panels (IKCScore, TNBC-ICI and DeepGeneX), and models that identifies significant genes from the testing cohorts (COMPASS, IRNet, NetBio, PRECISE and Tres). First each group, we firstly identify the number of overlaps between the combination of selections of the models. However, since different pathways may have similar expression and concepts, we input the pathways as one group and base on each dataset to EnrichmentMap[79] for further analysis, a bioinformatics tool that input relevant pathway information and build a network base on the overlapping genes between the pathways. For IRNet and NetBio specifically, because they do not provide significant genes. As a result, we ignored the edges strengths in the network and create a pseudo table for IRNet and NetBio that assumed all includes all genes inside the pathway have been involved in the dataset. AutoAnnotate was utilised to cluster the pathways, each with a concept. In this way we could study whether the models identify shared immunotherapy-related pathways and compare the performance in identifying important genesets for each modality. The full list of biological concepts, and presence of those concepts in each model is included in Supplementary Table S5.

## Author contributions

HAR and YG designed and conceptualized the study; YL wrote the paper. YL, HAR, YG, NF, MY, MH, LC, RA, AA, TP edited the manuscript. YL, LC carried out all the analyses, the supervision of HAR, AA. YL generated all figures and all tables. HAR, MY, MH, RA, LC are part of Global Research Impact Consortium (GRIC). The project has been designed under GRIC.

## Conflict of interest

The authors declare no competing financial and non-financial interests.

## Funding

This study was supported by the UNSW Scientia Program Fellowship and the Australian Research Council Discovery Early Career Researcher Award (DECRA), under grant DE220101210 to HAR. This collaborative project was also supported by the Second Century Fund (C2F), Chulalongkorn University. Additionally, the project has support from the Ratchadaphiseksomphot Endowment Fund, Top-up Grant of C2F, Chulalongkorn University.

## Data and Tool Availability

The code for preprocessing, running the models and evaluation can be accessed at https://github.com/jade0530/ICI_Prediction_Models_Benchmark_2025

# Figure Legends

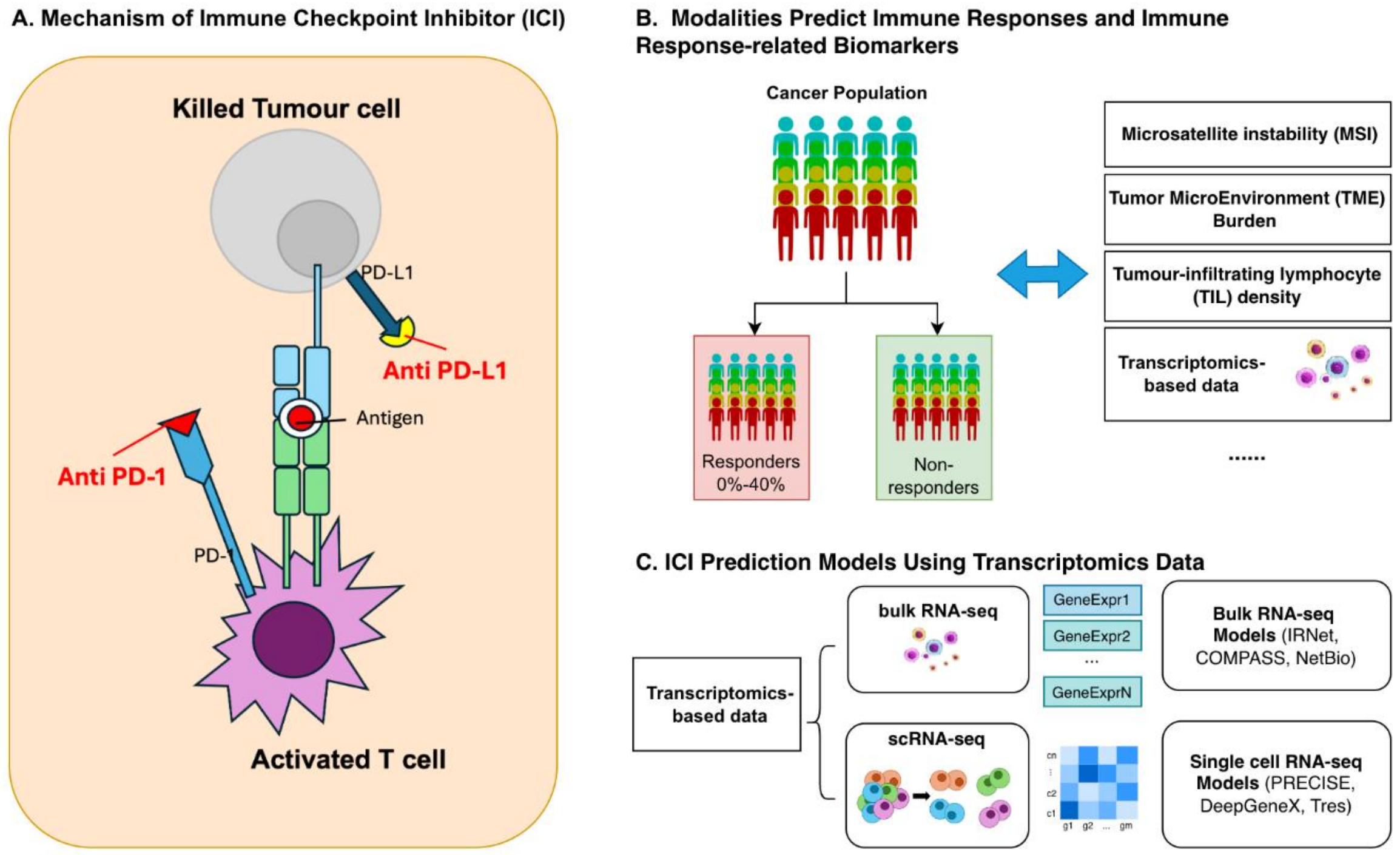

**Figure 1. Conceptual overview of immune checkpoint inhibition and transcriptomics-based ICI response prediction models. A.** Schematic of the mechanism of immune checkpoint inhibitors. Tumour cells expressing PD-L1 suppress T-cell activation through PD-1/PD-L1 signalling. Anti-PD-1 and anti-PD-L1 antibodies block this interaction, enabling activated cytotoxic T cells to eliminate tumour cells. **B.** Overview of clinical and molecular modalities used to predict ICI response. Across the cancer population, only 0-40% of patients achieve objective responses. Biomarkers such as microsatellite instability (MSI), tumour microenvironment (TME) burden, tumour-infiltrating lymphocyte (TIL) density, and transcriptomic signatures each capture different aspects of immune sensitivity. C) Transcriptomics-based prediction approaches used in this benchmark. Bulk RNA-seq provides sample-level gene expression profiles for models such as NetBio[35], IRNet[39], and COMPASS[38], while single-cell RNA-seq resolves cell-type-specific expression patterns underlying models including PRECISE[30], DeepGeneX[37], and Tres[32].

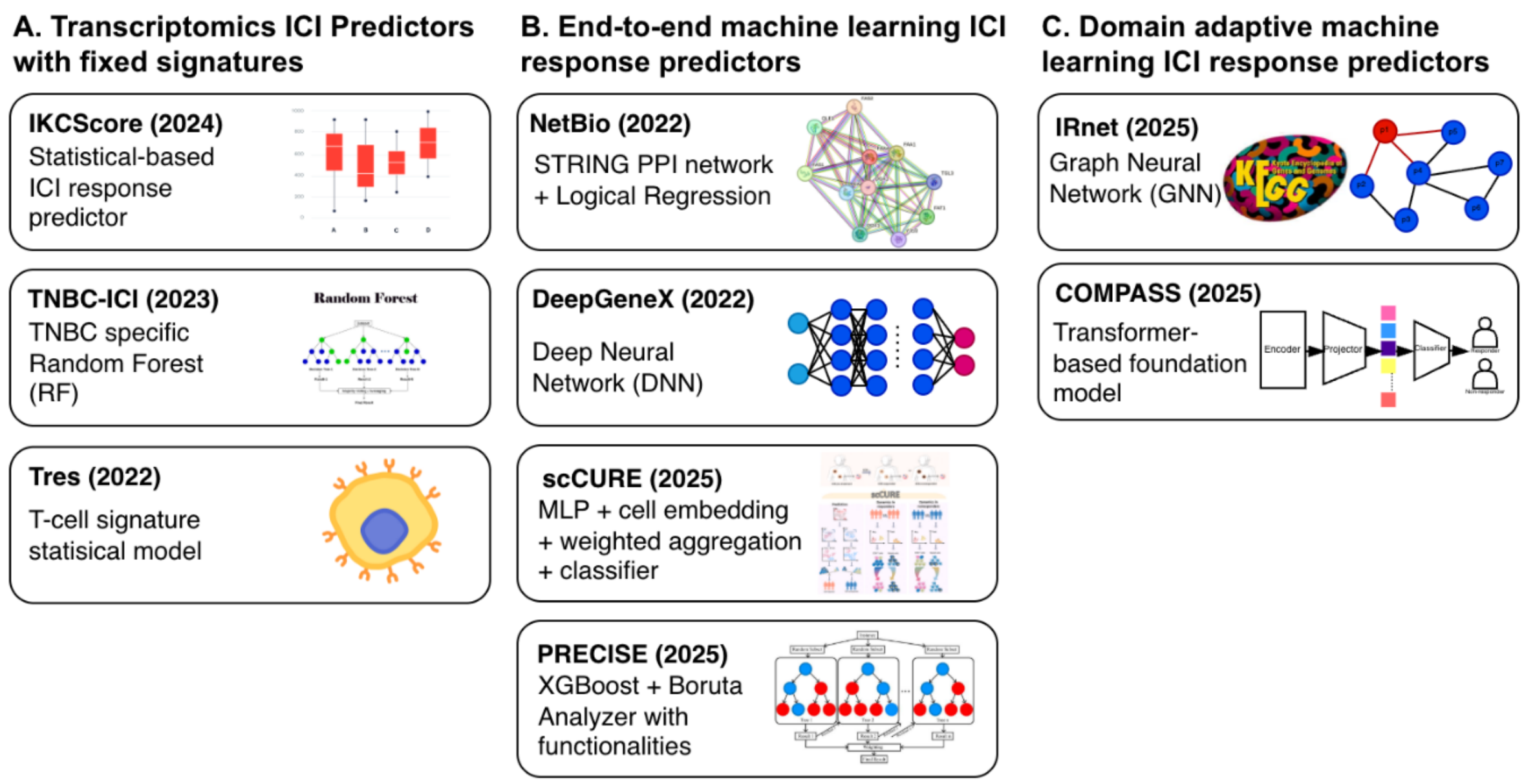

**Figure 2. Architectural categories of transcriptomics-based ICI response predictors.** Models are grouped into: **A.** fixed-signature statistical approaches (IKCScore, TNBC-ICI, Tres), **B.** end-to-end cohort-trained machine learning models (NetBio, DeepGeneX, scCURE, PRECISE), and **C.** domain-adaptive or pretrained frameworks (IRNet, COMPASS), illustrating differences in feature strategy and generalisation design.

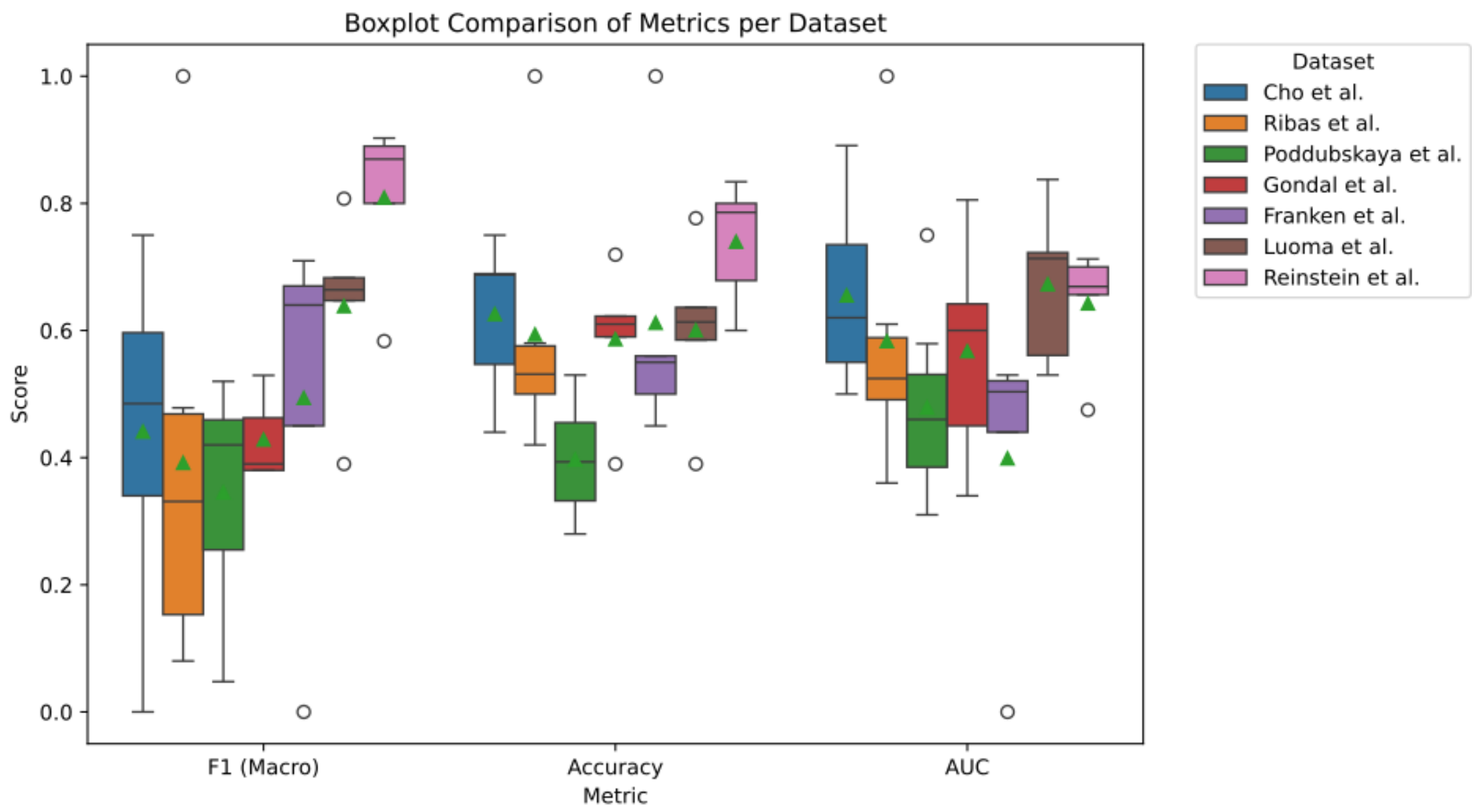

**Figure 3. Cross-dataset performance comparison of bulk and single-cell RNA-seq models.** Macro F1, accuracy, and AUC for all evaluated bulk RNA-seq models within bulk and single

cell RNA-seq dataset (Cho et al., Ribas et al., Poddubskaya et al., Gondal et al., Franken et al., Luoma et al., Reinstain et al.).

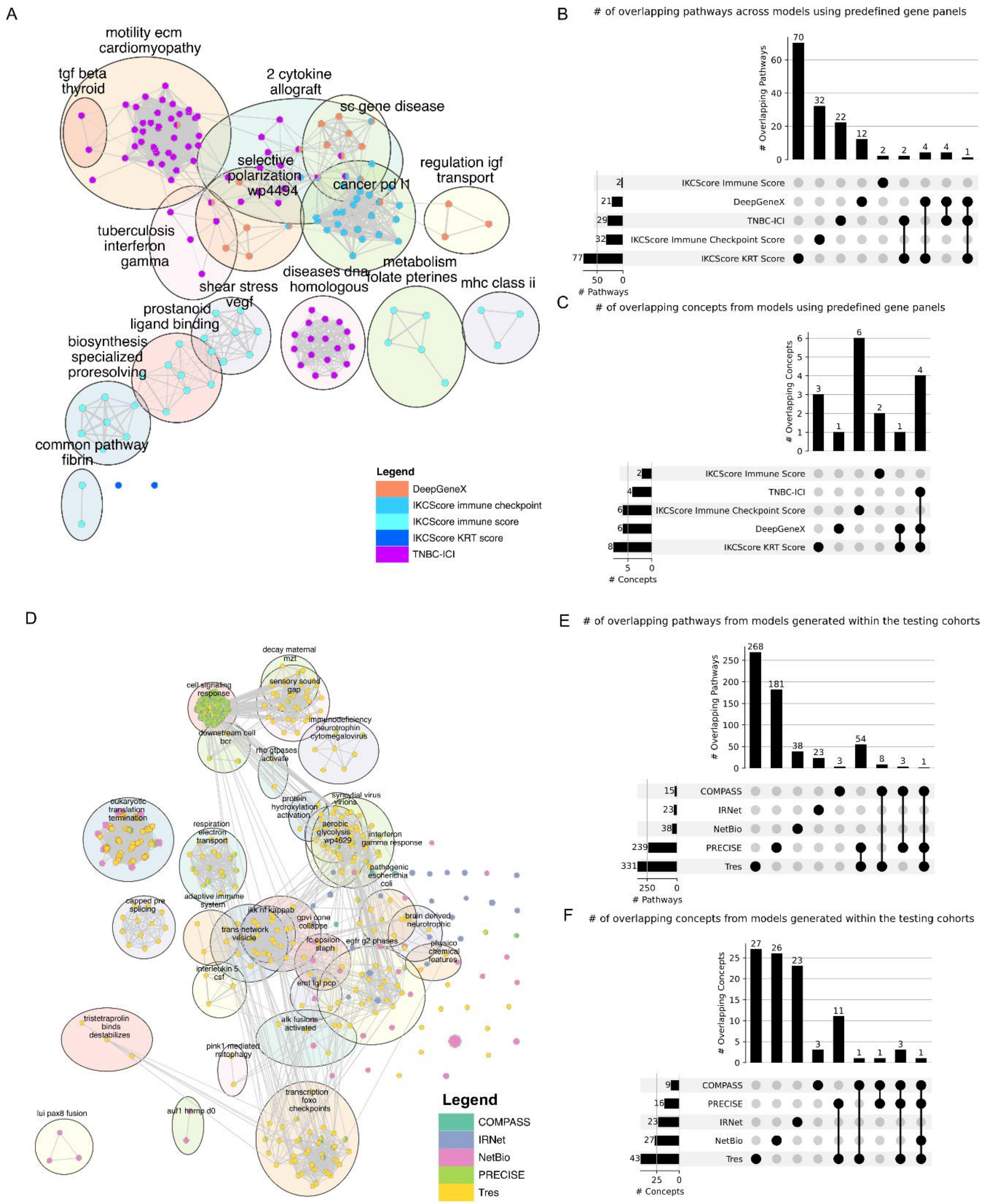

**Figure 4. Overlaps and interactions among significant biological pathways identified by each model across all datasets. A.** Pathway network clustered using EnrichmentMap pipeline for models utilising fixed gene panels (IKCScore, TNBC-ICI, Tres, DeepGeneX), with nodes representing pathways, edges representing gene-set overlaps, and each cluster indicating a

shared biological concept. **B.** UpSet plot for number of overlapping pathways across all models using fixed gene panels. **C.** UpSet plot showing the number of overlapping biological concepts (as defined in **A**) across models using fixed gene panels. **D.** Similar to **A**, Pathway network clustered using EnrichmentMap pipeline for models generate significant genes within the dataset, with nodes representing pathways, edges representing gene-set overlaps, and each cluster indicating a shared biological concept. **E.** UpSet plot for number of overlapping pathways across all models generate significant genes within testing cohorts. **F.** UpSet plot showing the number of overlapping biological concepts (as defined in **C**) across models generate significant genes within testing cohorts.

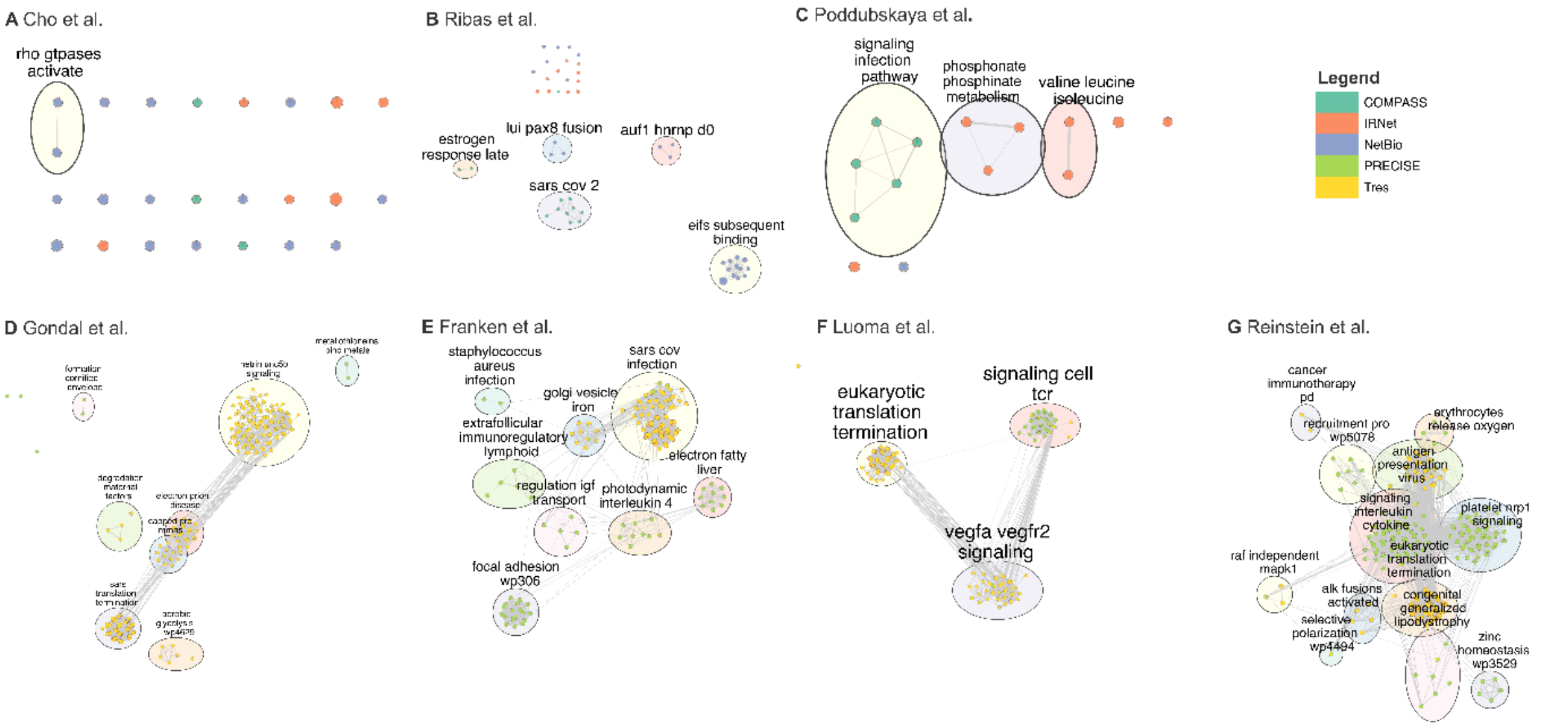

**Figure 5. Enrichment Map visualisation of pathway-level biological concepts across unseen cohorts.** EnrichmentMap networks depict biological concepts identified by pathway-level biomarkers for each dataset. Nodes represent enriched pathways, edges indicate gene-set overlap, and clusters denote groups of functionally related biological concepts. **A.** Cho et al.[64] (bulk RNA-seq), with node colours corresponding to bulk RNA-seq-based models (NetBio, COMPASS, and IRNet). **B.** Similar to **A**, but for Ribas et al.[45] **C.** Similar to **A**, but for Poddubskaya et al.[65] **D.** Gondal et al.[69] (scRNA-seq), with node colours corresponding to scRNA-seq-based models (Tres, and PRECISE). **E.** Similar to **D** but for Franken et al. [70] **F.** Similar to **D** but for Luoma et al.[71] **G.** Similar to **D** but for Reinstein et al.[72].

## Tables

**Table 1. Overview of training data, computational architectures, and reported predictive tasks for the transcriptomic ICI response models evaluated in this study.**

| Model | Pretraining Data and Normalisation Methods | Architecture / Methodology | Report Tasks |
|---|---|---|---|
| **NetBio (2022)** | Bulk RNA-seq | Network based (PPI network, biological pathways) + Logistic Regression (LR) | Predict ICI response; identify significant gene sets |
| **Tres (2022)** | scRNA-seq plus bulk T-cell | Signature-based statistical model | Predict ICI response; identify T-cell related resilient markers |
| **DeepGeneX (2022)** | Single-cell RNA-seq | Deep neural network (DNN) + feature elimination (multilayer DNN) | Predict ICI response, identify biomarker genes through feature elimination processes |
| **scCURE (2023)** | single-cell RNA-seq data (pan-cancer) | MLP + cell embedding + weighted aggregation + classifier | DEG; ICI response prediction; cell type changes identification |
| **IKCScore (2024)** | Bulk RNA-seq | Statistical model | IKCscore correspond to "hot/cold tumours" with higher score indicates higher possibility to be responsive |
| **TNBC-ICI (2024)** | Bulk RNA-seq (TNBC focused) | Gene panel selection + Random Forest | Predict ICI response from a pre-defined gene list |
| **COMPASS (2025)** | Bulk RNA-seq (pan-cancer) | Transformer-based model with interpretable layers | Predict ICI response; identify significant gene-level, gene set-level and cell-level features by concept projection |
| **IRnet (2025)** | Bulk RNA-seq (pan-cancer) | Graph Neural Network (GNN) | Predict ICI response; identify significant pathways |
| **PRECISE (2025)** | Single-cell RNA-seq (melanoma-focused) | XGBoost + Boruta Analyzer + Reinforcement Learning | Predict ICI response; Identify gene and cellular signatures of checkpoint response |

**Table 2 Performance of bulk and single-cell ICI response prediction models on six unseen cohorts, reported as accuracy, macro F1, AUC, class imbalance ratio, and runtime.** Bulk models were evaluated on bulk RNA-seq datasets (Cho et al.[77], Ribas et al.[49], Podduskaya et al.[78]), and scRNA-seq models on scRNA-seq datasets (Gondal et al.[82], Franken et al.[83], Luoma et al.[84], Reinstein et al.[85]).

| Dataset | Model | Accuracy | F1* (Macro) | AUC | Imbalance Ratio | Runtime (s) |
|---|---|---|---|---|---|---|
| Cho et al. | NetBio | **0.75** | **0.75** | 0.75 | 0.45 | 17.00 |
| | IKCScore | 0.69 | 0.62 | **0.89** | 0.45 | 1.85 |
| | TNBC-ICI | 0.69 | 0 | 0.5 | 0.45 | 1.66 |
| | IRnet | 0.69 | 0.54 | 0.69 | 0.45 | 6.95 |
| | COMPASS (LFT) | 0.44 | 0.31 | 0.55 | 0.45 | 24.60 |
| | COMPASS (PFT) | 0.50 | 0.43 | 0.55 | 0.45 | 21.43 |
| Ribas et al. | NetBio | 1.00 | 1.00 | 1.00 | 0.85 | 64.00 |
| | IKCScore | 0.5 | **0.48** | 0.52 | 0.85 | 1.39 |
| | TNBC-ICI | 0.56 | 0.22 | 0.52 | 0.85 | 1.65 |
| | IRnet | **0.58** | 0.44 | **0.61** | 0.85 | 12.04 |
| | COMPASS (LFT) | 0.42 | 0.13 | 0.36 | 0.85 | 25.44 |
| | COMPASS (PFT) | 0.50 | 0.08 | 0.48 | 0.85 | 23.70 |
| Podduskaya et al. | NetBio | **0.53** | 0.44 | 0.46 | 0.49 | 55.00 |
| | IKCScore | 0.39 | 0.48 | 0.58 | 0.49 | 2.29 |
| | TNBC-ICI | 0.34 | 0.05 | 0.48 | 0.49 | 1.76 |
| | IRnet (PD-L1) | 0.43 | 0.42 | **0.75** | 0.75 | 7.94 |
| | IRnet (PD-1) | 0.32 | 0.27 | 0.32 | 0.26 | 11.08 |
| | COMPASS (LFT) | 0.28 | 0.24 | 0.31 | 0.49 | 26.74 |
| | COMPASS (PFT) | 0.48 | **0.52** | 0.45 | 0.49 | 24.63 |
| Gondal et al. | DeepGeneX | 0.72 | 0.53 | 0.81 | 0.54 | 374.66 |
| | Tres | 0.39 | 0.38 | 0.34 | 0.54 | 0.89 |
| | scCURE | 0.62 | 0.46 | 0.64 | 0.54 | 263520.05 |
| | PRECISE | 0.59 | 0.38 | 0.45 | 0.54 | 595.02 |
| | PRECISE-refined features | 0.61 | 0.39 | 0.60 | 0.54 | 2905.85 |
| Franken et al. | DeepGeneX | 1.00 | 0.00 | 0.00 | 0.67 | 128.85 |
| | Tres | 0.45 | 0.45 | **0.53** | 0.67 | 0.51 |
| | PRECISE | 0.50 | 0.67 | 0.44 | 0.67 | 147.46 |
| | PRECISE-refined features | 0.55 | **0.71** | 0.52 | 0.67 | 811.20 |
| | scCURE | **0.56** | 0.64 | 0.50 | 0.67 | 200829.72 |
| Luoma et al. | DeepGeneX | 0.64 | 0.65 | 0.71 | 0.94 | 116.61 |
| | Tres | 0.39 | 0.39 | 0.53 | 0.94 | 0.53 |
| | PRECISE | **0.78** | **0.81** | **0.84** | 0.94 | 1238.37 |
| | PRECISE-refined features | 0.59 | 0.68 | 0.72 | 0.94 | 5297.96 |
| | scCURE | 0.61 | 0.66 | 0.56 | 0.94 | 226131.02 |
| Reinstein et al. | DeepGeneX | **0.80** | 0.89 | 0.70 | 0.25 | 58.77 |
| | Tres | 0.60 | 0.58 | 0.66 | 0.25 | 7.56 |
| | PRECISE | 0.68 | 0.80 | 0.48 | 0.25 | 1155.59 |

| | | | | | |
|---|---|---|---|---|---|
| PRECISE-refined features | 0.79 | 0.87 | **0.71** | 0.25 | 2958.51 |
| scCURE | 0.83 | **0.90** | 0.67 | 0.25 | 152123.31 |

**Table 3 Detailed information of unseen bulk RNA-seq cohorts used in the benchmark.**

| Cancer Type | Publication | # Samples | # responders | # non-responders | Accession ID |
|---|---|---|---|---|---|
| NSCLC | Cho et al. Nature 2020 | 16 | 5 | 11 | GSE126044 |
| Melanoma | Ribas et al., Nat Commun .2020 | 48 | 22 | 26 | GSE158403 |
| NSCLC | Poddubskaya et al., Immuno. 2024 | 61 | 41 | 20 | GSE274975 |

**Table 4 Detailed information of unseen scRNA-seq datasets used in this benchmark.**

| Cancer Type | Paper | Technology | # Samples | # pre responders | # pre-non-responders | # T cells pre-responders | # cells pre-non-responders | Accession ID |
|---|---|---|---|---|---|---|---|---|
| BCC, Melanoma, MBM, TNBC | Gondal et al. 2025 | 10X Genomics | 74 | 26 | 48 | 52291 | 77347 | https://zenodo.org/records/14511579 |
| HNSCC | Franken et al., Immunity 2024 | 10X Genomics | 20 | 8 | 12 | 61878 | 107316 | https://lambrechtslab.sites.vib.be/en/data-access |
| HNSCC | Luoma et al. 2022 | 10X Genomics | 33 | 17 | 16 | 14726 | 19880 | GSE200996 |
| | Reinstein et al. 2024 | 10X Genomics | 20 | 4 | 16 | 3743 | 22818 | GSE235090 |

## Supplementary Materials

**Supplementary Table S1.** Model inclusion-exclusion criteria. Models were included if they were published between May 2023 and November 2025, used scRNA-seq or bulk RNA-seq as input, were designed to predict ICI response, and had publicly available code. Models not meeting all criteria were excluded.

**Supplementary Table S2.** Metadata summary of transcriptomic datasets used for benchmarking. Overview of selected bulk and single-cell RNA-seq datasets included in this study. For each dataset, the table reports cancer type, source publication, reference identifiers (PMID), tissue of origin, dataset accession ID, sequencing technology (for scRNA-seq only), reference genome, number of genes, number of samples, and the distribution of pretreatment responders and non-responders. For scRNA-seq datasets, the number of responder and non-responder cells and cell types utilised in this study are additionally reported.

**Supplementary Table S3.** Biological pathways for models extracting geneset features, NetBio (2022) and IRNet (2025), with z-score >= 2.

**Supplementary Table S4.** EnrichR results of biological pathways for models extracting genes, Tres (2022), DeepGeneX(2022), TNBC-ICI (2023), IKCScore (2024), COMPASS (2025) and PRECISE (2025), with p-value <0.05.

**Supplementary Table S5.** Presence of enriched biological concepts across ICI prediction models that identifies important genes within the testing cohort. Binary matrix indicating the presence (1) or absence (0) of enriched biological concepts identified by each transcriptomics-based ICI prediction model.

**Supplementary Figure S1.** Annotation and Differentially Expressed Gene for Reinstein et al.